 \documentclass [12pt,a4paper      ]{article}
\usepackage{graphics}
\usepackage{times}

\DeclareFontFamily{OT1}{times}{}
\DeclareFontShape {OT1}{times}{m }{n }{ <-> ptmr }{}
\DeclareFontShape {OT1}{times}{bx}{n }{ <-> ptmb }{}
\DeclareFontShape {OT1}{times}{m }{it}{ <-> ptmri}{}
\DeclareFontShape {OT1}{times}{bx}{it}{ <-> ptmbi}{}
\usepackage{amsmath}
\usepackage{amsfonts}
\usepackage{amssymb}
\usepackage{latexsym}
\numberwithin{equation}{section}

\begin{document}

\title{{\bf \vspace{-2.5cm} Antimatter underestimated} \footnote{Published in Nature {\bf 325} (26 February 1987) 754; reprinted in Bulletin of Peace Proposals {\bf 19} (1988) 459--460.  Some minor errors and changes in emphasis made by the \emph{Nature} editorial staff have been corrected to conform to the originally submitted letter.}}

\author{{\bf Andre Gsponer and Jean-Pierre Hurni}\\
\emph{Independent Scientific Research Institute}\\ 
\emph{Box 30, CH-1211 Geneva-12, Switzerland}\\
e-mail: isri@vtx.ch\\}

\date{Version ISRI-87-01.2 ~~ \today}

\maketitle

\begin{abstract}

We warn of the potential nuclear proliferation's consequences of military applications of nano- or microgram amounts of antimatter, such as triggering of high-yield thermonuclear explosions,  laser pumping, compact sources of energy, directed-energy beams, and portable sources of muons.

\end{abstract}

\noindent Sir --- The  first  published account of success  in  immobilizing antiprotons in an electromagnetic trap has been the subject of  a recent leading article News  and Views editorial  \cite{1}.   Although this  article gives a good overview of the scientific meaning  of this  historical  achievement,  it  gives in our  view  a  rather misleading perspective on its likely military significance \cite{2}. 
 
     It is true that for some applications of antimatter, such as antiballistic missile or spacecraft propulsion,  relatively large amounts of antiprotons are required.   For most other cases,  the useful amounts are usually much smaller.   For example,  we have calculated that  less  than  a  microgram  of  antiprotons   is sufficient  to  trigger  a  thermonuclear  explosion  or  pump  a powerful X-ray laser \cite{3,4}. 
 
     But  antimatter  is not only the most powerful of all  high-explosives,  it  is  also the only feasible  portable  source  of muons.   In every antiproton annihilation,   on the average three muons are produced.  These could be used to induce muon-catalysed fusion  reactions in a deuterium-helium-3 mixture,  an attractive solution for a low-weight neutron-free space nuclear reactor that could be operated in a continuous or pulsed mode. 
 
     Furthermore,  by collecting and cooling the muons (a relatively easy  task  compared  with that of cooling  antiprotons \cite{5})  a  very intense  beam  could  be formed and sent into the  atmosphere  to guide,  over  a range of more than 10 km,  a series of  powerful electron  or proton beam pulses towards a target.   More  simply, stopping  the  muons  in a suitable material  would  generate  an extraordinarily  effective  X-ray  lasing  medium,  for  the  two microseconds lifetime of the muonic-atoms. 
 
     In  outer space,  a very low intensity burst of  antiprotons would  be most suitable for active warhead/decoy  discrimination. In  this  and  the two previous examples,  the needed  amount  of antiprotons  is  of  the  order  of  \emph{nanograms}  per   engagement. Conservative  estimates  of  the technical problems  involved  in producing  and manipulating \emph{microgram} amounts of  antimatter  per day,  show  that  known technology is only a couple of orders  of magnitudes away from meeting the challenge \cite{4,6,7}. 
 
     We   are  very  much  concerned  by  the  implications for nuclear   weapons proliferation   of  the   undisputable   scientific feasibility  of several antimatter weapon concepts.   To  develop such  weapons would add considerable impetus to the current  arms race.   We  call  for an immediate ban on all antimatter  related research, especially since this work is fundamental to many fourth-generation nuclear weapon systems.


\begin{thebibliography}{999}

\bibitem{1} J. Maddox, \emph{How to make antimatter last}, Nature {\bf 324} (27 November 1986) 299.
 
\bibitem{2} A. Gsponer et J.-P. Hurni, \emph{Les armes \`a antimati\`ere},  La Recherche {\bf 17} (novembre 1986) 1440--1443. English translation: \emph{Antimatter weapons}, Bulletin of Peace Proposals {\bf 19} (1988) 444--450, e-print arXiv:physics/0507132 available in PDF format at\\ \underline{http://arXiv.org/pdf/physics/0507132}

\bibitem{3} A. Gsponer and J.-P. Hurni, \emph{The physics of antimatter induced fusion and thermonuclear explosions}, in G. Velarde and E. Minguez, eds.,  Proceedings of the 4th International Conference on Emerging Nuclear Energy Systems, Madrid, June 30/July 4, 1986 (World Scientific, Singapore, 1987) 166--169,  e-print arXiv:physics/0507114 available in PDF format at\\ \underline{http://arXiv.org/pdf/physics/0507114}
 
\bibitem{4} A. Gsponer and J.-P. Hurni, \emph{Antimatter 
induced  fusion and thermonuclear  explosions}, Atomkernenergie $\cdot$ Kerntechnik (Independent Journal on Energy Systems and Radiation) {\bf 49} (1987) 198--203,  e-print arXiv:physics/0507125 available in PDF format at\\ \underline{http://arXiv.org/pdf/physics/0507125} 

\bibitem{5} D. Neuffer, \emph{Principles and applications of muon cooling}, Fermi National Laboratory report FN-378 (January 1983) 31~pages, published in Part. Accel. {\bf 14} (1983) 75.

\bibitem{6} B.W. Augenstein: \emph{Concepts, problems, and 
opportunities for use of annihilation  energy,  
Prepared for the United States Air Force}, RAND 
Note N--2302--AF/RC (June 1985) 61~pages.

\bibitem{7}  R. Walgate, \emph{Defence lobby eyes antimatter}, Nature {\bf 322} (21 August 1986) 678.


\end{thebibliography}
\end{document}